\documentclass[conference]{IEEEtran}
\IEEEoverridecommandlockouts

\usepackage{cite}
\usepackage{amsmath,amssymb,amsfonts}
\usepackage{graphicx}
\usepackage{textcomp}
\usepackage{xcolor}
\usepackage{booktabs}
\usepackage{subcaption}
\usepackage{makecell}
\usepackage{multirow}
\usepackage[table]{xcolor}
\usepackage[ruled,vlined]{algorithm2e}
\usepackage[section]{placeins}
\def\BibTeX{{\rm B\kern-.05em{\sc i\kern-.025em b}\kern-.08em
    T\kern-.1667em\lower.7ex\hbox{E}\kern-.125emX}}

\begin{document}

\title{Foreground Voice Activity Detection: Learning Speaker Selectivity from Supervision\\
}

\author{
\IEEEauthorblockN{Guangzhao Yang\textsuperscript{*},
Muhammad Huzaifah\textsuperscript{*},
Yu Pan,
Jinya Sakurai,
Ningjie Bai\textsuperscript{\dag}}
\IEEEauthorblockA{R\&D Team, Recho Inc., Tokyo, Japan}
\texttt{\{g.yang, m.huzaifah, y.pan, j.sakurai, n.haku\}@recho-ai.com}
\thanks{\textsuperscript{*}Equal contribution.}
\thanks{\textsuperscript{\dag}Corresponding author}
}

\maketitle

\begin{abstract}
Voice activity detection (VAD) fronts most voice-agent pipelines, yet production detectors treat all human speech, background talkers included, as valid activity; in crowded settings this floods recognition, stalls turn-taking, and triggers false barge-in. We formalize Foreground VAD (FVAD): a frame-synchronous, enrollment-free task in which only the dominant speaker, defined by sustained presence rather than instantaneous loudness, is positive, and which reduces to conventional VAD when a single speaker is present. We show that foreground selectivity is largely governed by training supervision: the crucial ingredient is an augmentation recipe pairing foreground-only labels with competing-speaker mixing, generated fully automatically without human annotation. To quantify selectivity we introduce the Background False-Alarm Rate (BG-FAR), gated by foreground F1, and build a controlled benchmark, Mix-Interference, complemented by an adapted VOiCES for real-world far-field evaluation. Across equal-size backbones, Mamba and LSTM perform on par while a longer-context attention model is no better, suggesting that training supervision plays a substantially larger role than temporal modeling capacity in achieving foreground selectivity. The resulting lightweight streaming model, Mamba-FVAD, outperforms commercial VADs and enrollment-based speaker-aware systems in foreground selectivity while staying competitive on conventional VAD, at 1--2\,ms per-frame CPU latency.


\end{abstract}


\section{Introduction}
\label{sec:intro}


The rapid adoption of voice-based AI agents in customer service, personal assistants, and human–AI collaboration has raised the demand for natural, reliable spoken interaction. Voice activity detection (VAD), which decides whether each audio frame contains speech, is typically the first stage of such pipelines, ahead of automatic speech recognition (ASR), and its accuracy strongly shapes overall system performance.

Modern VAD systems~\cite{SileroVAD,TENVAD,zhang2018deep} achieve strong robustness, ultra-low latency, and efficient CPU-only inference at 30\,Hz or higher, yet share a basic limitation: they classify all human speech---background speakers included---as valid activity. Methods that do distinguish speakers supply an explicit prior in the form of an enrolled embedding, as in personal~\cite{ding2020personalvadspeakerconditionedvoice,ding2022personalvad20optimizing} and target-speaker VAD~\cite{wang2022targetspeakervoiceactivity}, rather than expecting a plain VAD to learn it. One might assume the limitation is architectural: deployed VADs often use LSTMs or GRUs~\cite{gudepu2023dernn,SHARMA2022116} with only a few hundred milliseconds of effective context, so isolating a foreground speaker would seem to need much longer-range modeling. In this paper, we test this directly: is the inability to separate foreground from background speech caused by the limited context of recurrent architectures, or by training data and objectives that never require the distinction? Our results strongly support the latter.

The limitation becomes severe in crowded settings such as restaurants, public squares, and meeting rooms, where background conversation and babble are misclassified as valid speech. For a voice agent this causes three failures: (1)~Irrelevant speech reaches downstream ASR, which modern far-field and noise-robust models~\cite{shi2026qwen3asrtechnicalreport,rouvier-mohammadamini-2022-far} transcribe into unrelated content resulting in erroneous LLM responses; (2)~Persistent background speech withholds the VAD falling edge that triggers generation, leaving the agent in a perpetual listening state; (3)~Background talkers falsely trigger barge-in\cite{chen2025fireredchatpluggablefullduplexvoice}, degrading the user experience.


We formalize the underlying task as \textbf{Foreground Voice Activity Detection (FVAD)}: a frame-synchronous binary classification in which a frame is positive if and only if it contains speech from the foreground speaker of the interaction. The foreground role is defined by sustained presence and temporal identity coherence, not instantaneous energy; the foreground speaker is the one with the greatest sustained speech presence, and a momentarily louder competitor must not capture the role (defining it by frame-wise loudness would otherwise collapse the task into energy-based VAD). Non-foreground speech is therefore negative, and with a single speaker FVAD reduces to conventional VAD. Formally, for a stream $x$ with per-frame labels $y_\tau$, personal VAD (pVAD) models $P(y_\tau \mid x, e_t)$ over three classes, target-speaker speech (tss), non-target-speaker speech (ntss), and non-speech (ns), given an exogenous enrollment $e_t$ of a known target. Instead, FVAD models the binary posterior $P(y_\tau{=}1 \mid x)$ for the foreground speaker $F(x)$, an endogenous function of the signal inferred online; a self-derived pseudo-enrollment the model forms from the audio and keeps coherent over time. Equivalently, FVAD's positive class is the analogue of tss with a self-inferred target, while ns and ntss collapse into its negative class. In short, pVAD is told who the target is, but FVAD must decide for itself (we assume one foreground speaker per segment, the identity the model commits to and tracks throughout). These choices distinguish FVAD from existing paradigms on three axes, expanded in Section~\ref{sec:related}: it strictly generalizes VAD (no separation front-end), it is enrollment-free yet identity-committed (unlike pVAD or energy-following detection), and it runs in a single streaming stage (unlike a VAD-plus-diarization cascade).

Our central finding is that foreground focus is governed largely by \emph{supervision}, not architecture. The decisive factor is the training-data recipe: clean utterances are pseudo-labeled by a of state-of-the-art VAD, then unlabeled interfering speakers and babble are mixed into the target at controlled signal-to-noise ratios (SNRs), level dynamics, and far-field reverberation, with the labels kept on the foreground alone. The model is thereby supervised to treat competing speech as negative, and a sustained foreground identity is encoded implicitly by the supervision. Because selectivity is conferred by the data, the backbone is free to be chosen for deployment. Our primary system, \emph{Mamba-FVAD}, pairs a learnable LEAF front-end~\cite{zeghidour2021leaflearnablefrontendaudio} with a Mamba state-space backbone~\cite{gu2024mambalineartimesequencemodeling} whose linear-time recurrence gives constant per-frame computation and memory for low-latency, CPU-only operation. Varying the backbone confirms architecture is not the bottleneck: an LSTM~\cite{hochreiter1997lstm} (a recurrent model with more limited context~\cite{11230983}) matches it, while a long-context attention model~\cite{vaswani2023attentionneed} yields no meaningful gain.

In summary, our key contributions are as follows:
\begin{itemize}
\item \textbf{Task and metric.} We give a formal definition of enrollment-free foreground (dominant-speaker-selective) VAD, and introduce the Background False-Alarm Rate (BG-FAR) metric: the probability of firing while only competing speech is active and the foreground is silent, gated by foreground F1, directly measuring the identity-commitment the task demands.
\item \textbf{New benchmark.} We introduce Mix-Interference, in which the target foreground track is mixed with competing speech at controlled SNRs, and adapt VOiCES~\cite{richey18_interspeech} (real far-field television and babble noise) as a cross-check. We will release Mix-Interference to spur further work on FVAD.
\item \textbf{Data recipe.} We propose a fully automatic pipeline that yields large-scale frame-level FVAD supervision without human annotation, and show through controlled architecture and data ablations that the recipe is what primarily unlocks foreground focus.
\item \textbf{State-of-the-art model.} We develop \emph{Mamba-FVAD}, a lightweight streaming model that surpasses mainstream commercial VADs and specialized speaker-aware systems on the proposed benchmark while remaining competitive on conventional VAD in clean and noisy conditions.
\end{itemize}

\section{Related Work}
\label{sec:related}
Production VADs, including Silero~\cite{SileroVAD}, TEN-VAD~\cite{TENVAD}, MarbleNet~\cite{jia2021marblenetdeep1dtimechannel}, and related neural detectors~\cite{zhang2018deep,gudepu2023dernn,SHARMA2022116}, give robust, low-latency frame-level detection but are trained to detect \emph{all} speech. For example, competing speakers are treated as positive examples rather than interference, so speaker selection is therefore not naturally carried out---a gap we address with FVAD. A common deployment remedy gates a VAD by signal energy, assuming the foreground is closest to the microphone; but instantaneous energy is an unreliable proxy for the foreground role. For instance, foreground speech may be soft at utterance boundaries, whereas a background interjection can be momentarily loud, conditions our benchmark probes. Furthermore, smoothing energy statistics over a longer analysis window trades flicker for latency, and the gate's hand-tuned thresholds transfer poorly across devices and environments. Speech enhancement front-ends~\cite{zhao2025clearervoicestudiobridgingadvancedspeech,defossez2020realtimespeechenhancement} raise SNR before VAD, but optimize perceptual quality over speaker selectivity. Likewise, they are trained to preserve all speech; they denoise background speakers alongside the foreground and erase the very level and reverberation cues that distinguish them, while adding cascade latency in the single-speaker conditions that dominate real usage. By contrast, FVAD is a \emph{strict generalization of VAD}, reducing exactly to conventional detection when one speaker is present and requiring no separation front-end.

A second line incorporates speaker identity. Streaming end-to-end diarization~\cite{liang2023framewisestreamingendtoendspeaker,liang2025lseendlongformstreamingendtoend} attributes all speakers (``who spoke when'') and typically needs look-ahead, leaving the foreground unchosen. Personal and target-speaker VAD~\cite{ding2020personalvadspeakerconditionedvoice,ding2022personalvad20optimizing,wang2022targetspeakervoiceactivity} condition on an enrollment embedding, giving strong selectivity only for a pre-registered identity and failing when the speaker is unknown or enrollment is mismatched. Post-VAD diarization~\cite{bredin2019pyannoteaudioneuralbuildingblocks} is accurate offline but introduces multi-stage latency. FVAD differs from previous paradigms in several ways: the foreground is inferred endogenously from the signal (no enrollment), committed to as a single coherent identity (unlike diarization, which tracks everyone, or energy- or saliency-following gates, which follow whoever is momentarily loudest), and produced by one streaming detector (no cascade).

\section{Methodology}
\subsection{Interference-Aware Data Recipe}
\label{sec:IAdata}
\textbf{Training data.} We propose a fully automatic pipeline for constructing large-scale VAD training data without human annotation. Clean utterances are first pseudo-labelled by Silero-v6~\cite{SileroVAD}, then passed through an energy-adaptive edge refinement that relocates each segment's onset and offset to the local acoustic boundary. This places the decision threshold a fixed fraction of the way from the local noise floor toward speech energy, with separate onset/offset factors to absorb the asymmetric boundary lag of the base detector. The refinement algorithm in effect tightens the labels without discarding low-energy speech. The training corpus was assembled from an in-house Japanese call-centre dataset of real conversational speech (1,128 hours) and concatenated LibriSpeech read speech with inter-utterance silences, similar to the recipe used by LibriVAD~\cite{stylianou2025librivadscalableopendataset}.

\textbf{Dynamic Augmentation.} Rather than pre-computing fixed mixtures, augmentations are applied stochastically per utterance at load time, so every epoch presents new conditions. The selectivity-bearing component is competing-speaker mixing. One to three interfering-speaker segments, sampled from a held-out speech pool and covering 10–50\% of the utterance, are mixed into the foreground at a target-to-interference ratio (measured over speech-active frames) of 0–15 dB, with a probability p of 0.2. Each interfering segment is rendered as a background source via far-field simulation by convolving it with a room impulse response (RIR) sampled from the DNS-Challenge~\cite{dubey2023icassp} (50/50 simulated and real recorded RIRs) and applying a random 1–4 kHz low-pass that models far-field high-frequency roll-off. The foreground labels are left unchanged, so this competing speech is supervised as negative. The remaining augmentations target general robustness, each applied independently: environmental-noise mixing (p$=$0.45; scattered or full-length; bimodal SNR, easy [5, 20] dB / hard [-5, 5] dB), music mixing (p$=$0.15; [5, 20] dB), room reverberation on the target (p$=$0.2), gain perturbation ($\pm$6 dB) with per-segment level dynamics, hard non-speech negatives (noise-, silence-, music- and far-field-speech-only clips, 1–2\% each), and telephone band-pass filtering (300–3400 Hz / 50–7000 Hz), sharpening rejection without eroding foreground recall.

\subsection{Model Architecture and Experimental Setup}

\begin{figure}[h]
    \centering
    \includegraphics[width=0.49\textwidth]{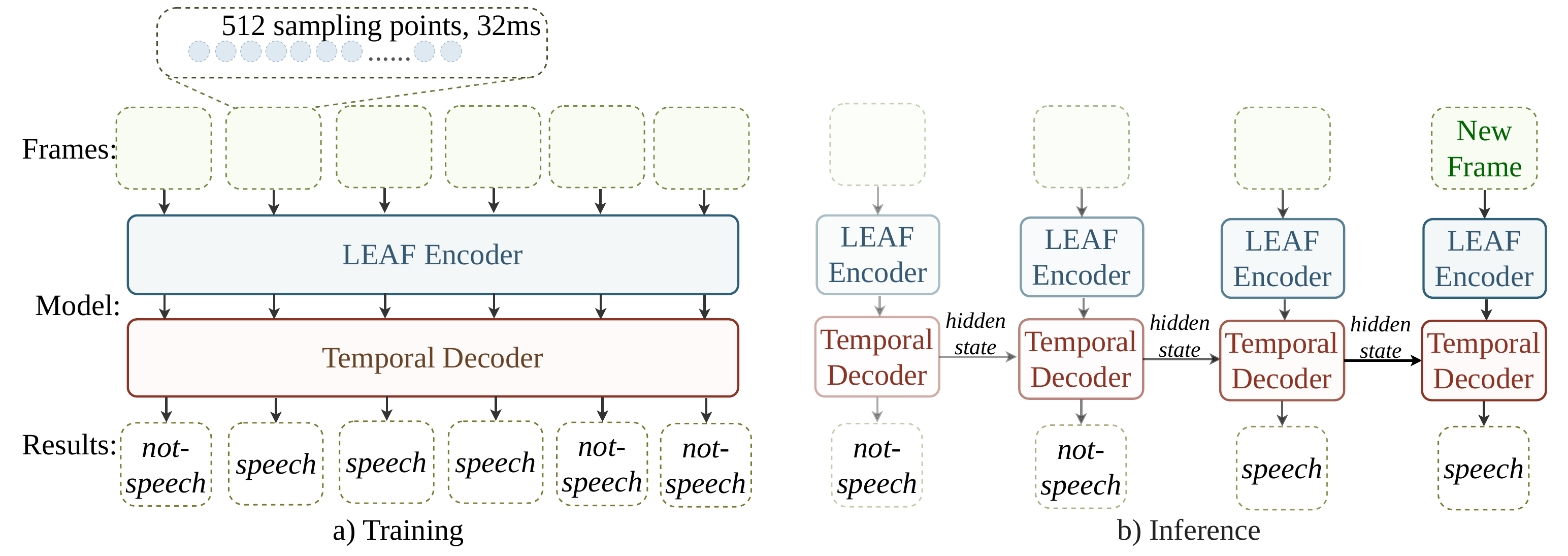}
    \caption{Overall architecture of the proposed Mamba-FVAD framework.}
    \label{fig:architecture}
\end{figure}

To train for FVAD, we pair a learnable acoustic front-end with a temporal backbone and a per-frame classification head, processing raw audio frame-wise (Fig.\ref{fig:architecture}). The front-end is a LEAF encoder\cite{zeghidour2021leaflearnablefrontendaudio} adapted for streaming: temporal pooling is removed and Gaussian pooling replaced by global average pooling to preserve frame-level resolution, operating on pre-segmented 512-sample frames at a fixed 31.25\,Hz. Because selectivity is conferred by the data (Sec.\ref{sec:IAdata}), the backbone is free to be chosen for deployment. Our primary system, \emph{Mamba-FVAD} ($\sim$0.6M parameters), uses a Mamba state-space decoder~\cite{gu2024mambalineartimesequencemodeling}, whose linear-time recurrence gives $\mathcal{O}(1)$ per-frame compute and memory and thus low-latency, CPU-only real-time operation. To study how much temporal context foreground selectivity requires, we compare three iso-parameter backbones spanning context capacity under an identical recipe: the Mamba above; an \emph{LSTM}~\cite{hochreiter1997lstm}, that is a recurrent model with more limited effective context~\cite{11230983}; and a \emph{Transformer}~\cite{vaswani2023attentionneed}, with unbounded in-window context but length-growing cost, serving as an offline upper bound. Mamba and the LSTM run at $\mathcal{O}(1)$ per-frame cost and stream; the Transformer does not.

All backbones train with AdamW (lr $10^{-4}$, weight decay $0.01$, cosine schedule with linear warmup), batch size 16, dropout 0.3, and bf16 mixed precision; and the Transformer a reduced learning rate and longer warmup, for stable convergence. We call models trained with our recipe (Sec.~\ref{sec:IAdata}) \emph{interference-aware} (IA). As a control, we also train each backbone under the conventional LibriVAD recipe~\cite{stylianou2025librivadscalableopendataset} to isolate the recipe's contribution.

\subsection{Benchmark Creation}
\textbf{Mix-Interference} is a controlled benchmark that pits a single foreground speaker against one competing background speaker at known relative levels. Foreground utterances are drawn from concatenated LibriSpeech~\cite{librispeech} test-other recordings, with frame-level VAD labels (31.25 Hz) from an energy-based labeler, where these labels define the foreground (target) throughout. For each foreground clip we synthesize seven temporally-aligned 16 kHz variants: (1) the clean target; (2) target + environmental noise, and (3–7) target + the same noise + a competing speaker at five fixed target-to-interference SNRs of {9, 11, 13, 15, 17} dB. The competing speaker is a different test-other utterance, passed through a far-field simulation (random small/medium room, near/medium distance, 4–8 kHz low-pass filter) and offset by a random temporal delay ($\pm$10 s) so it behaves as an intermittent background talker. All level ratios are computed over speech-active frames (VAD-weighted RMS): noise is mixed at $\approx$15 dB SNR for the interference variants and a harsher $\approx$6 dB for the noise-only variant, and each mixture is peak-normalized to 0.95. Crucially, every variant retains the foreground-only VAD labels; the competing speaker is never labeled as speech so the benchmark directly measures whether a model tracks the foreground while rejecting background speech. Because the five mixed variants share the same target, noise, and interferer and differ only in the interferer's level, the takes are frame-locked, enabling paired per-frame contrasts.

\textbf{VOiCES} devkit~\cite{richey18_interspeech} was adapted as an out-of-distribution FVAD benchmark to test whether foreground selectivity transfers to physically recorded far-field speech. Unlike Mix-Interference, nothing is mixed at conversion time: clean LibriSpeech utterances are replayed through a loudspeaker and re-recorded by far-field microphones in real rooms, so reverberation, distance, and the distractor, categorized as none (clean), musi (music), or babb/tele (competing background speech), are physically baked into the signal. We keep the distant recording as input and attach foreground-only labels by keying the target speaker's word-level timestamps from the original source forced alignments to each segment, while the recorded background talkers in babb/tele stay unlabeled, i.e. negative. This makes the competitors real recorded speakers rather than synthetic mixtures. Since every segment is recorded under all conditions and frame-aligned, the matched none take serves as a per-frame clean reference for paired metrics like BG-FAR. 

\subsection{Evaluation Metrics}
Conventional VAD scores, namely speech against non-speech F1 or ROC-AUC, cannot measure FVAD because they treat all speech as positive; a perfect foreground detector that suppresses a competing speaker would be charged with false negatives. We therefore evaluate against foreground labels with two complementary frame-level metrics that separate the two distinct ways an FVAD model can fail, either by missing the target, or firing on a competitor.

\textbf{Foreground F1} measures how well the model tracks the target. With foreground-speech frames as positives, it is the harmonic mean of precision and recall. Recall penalizes missed foreground speech and precision penalizes activation on any foreground-silent frame (silence, noise, or competing speech). Reported together with its recall component, it acts as a gate; a model cannot appear ``selective" merely by being conservative as suppressing the target collapses recall and hence F1.

\textbf{Background False-Alarm Rate (BG-FAR)} isolates the behavior that defines the task, that is, firing on a competing background speaker. It is the false-alarm rate computed only over frames where the foreground is silent yet a background source is active: 
\[\text{BG-FAR} = P(y_\tau=1 \mid \text{foreground-silent} \wedge \text{background-active}) \] 

Ordinary FAR over all silent frames is dominated by easy true silence and dilutes this signal, by conditioning on background-active frames, we target exactly the hard case. We obtain a frame-exact background-active mask without extra annotation by exploiting paired, frame-aligned takes from the benchmark datasets. Each clip has a matched reference with the same foreground and environment but no competing speaker (the noise-only take in Mix-Interference, the none take in VOiCES), and a foreground-silent frame is marked background-active when its energy exceeds the reference by more than a margin $\delta$ (6 dB in this paper). BG-FAR is bounded by 0 and 1, denoting complete rejection and complete acceptance respectively. In our evaluation, BG-FAR is calculated at an operating threshold of 0.5.

The two metrics are decisive only in combination. Genuine foreground selectivity requires both high Foreground F1 (the target is detected and the model does not over-fire) and low BG-FAR (competitors are rejected). Neither suffices alone, for example, a silent model trivially attains BG-FAR $=0$ but fails Foreground F1, whereas a generic VAD attains high recall but high BG-FAR. The full algorithms are described in Appendix A and B.

\section{Results and Analysis}

\subsection{Foreground VAD Performance}
\label{sec:bgfar}

\begin{figure}[htbp]
    \centering
    \includegraphics[width=0.90\columnwidth]{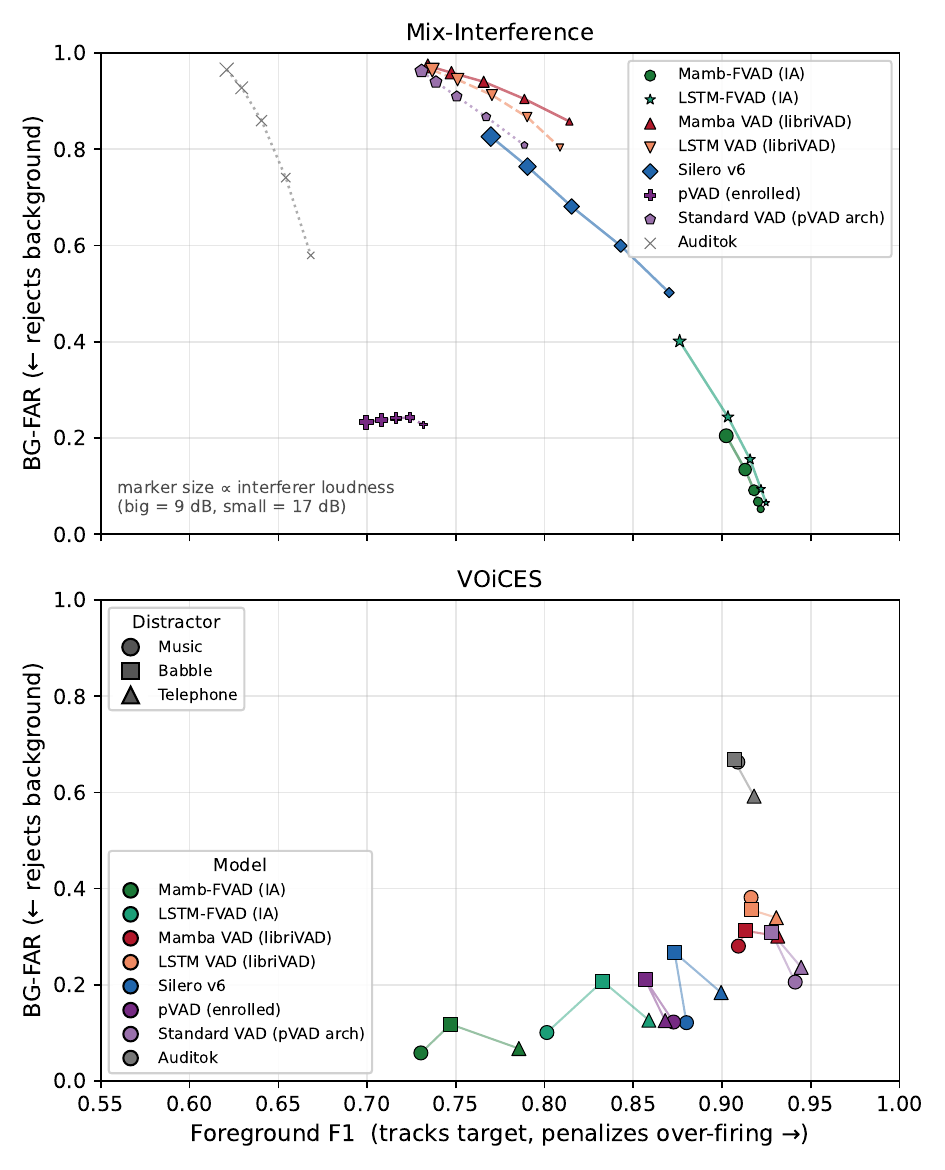} 
    \caption{BG-FAR vs. Foreground F1 for Mix-Interference swept over SNR (9--17\,dB) (top), and VOiCES by distractor (bottom). The ideal FVAD model falls in the bottom-right.}
    \label{fig:bgfar_stack}
\end{figure}


Fig.~\ref{fig:bgfar_stack} plots BG-FAR against Foreground F1 for our IA- and LibriVAD-recipe models against baselines: production-grade Silero-v6, a pretrained pVAD with speaker enrollment$^1$, a Standard VAD (pVAD's architecture, no enrollment), and energy-based Auditok. An ideal FVAD model sits bottom-right with high foreground tracking, low BG-FAR. On Mix-Interference our IA models (Mamba-FVAD and its iso-parameter LSTM) occupy this region (Foreground F1 0.88-0.92; BG-FAR from 0.05 at 17\,dB to at most 0.40 at 9\,dB), while every other system drifts up the BG-FAR axis as the interferer becomes louder, exceeding 0.8 at the loudest. Critically, the LibriVAD-recipe models share the same architectures yet collapse into this high-BG-FAR region, isolating the data recipe as the cause. Among baselines, only the enrolled pVAD keeps BG-FAR low at a flat ${\sim}0.23$ band, interferer-independent (as expected from an external speaker prior) but at far lower Foreground F1. Removing that prior (Standard VAD) returns BG-FAR to the LibriVAD region.

{
\let\thefootnote\relax
\footnote{$^1$https://github.com/pirxus/personalVAD}
}


\begin{table*}[t]
\centering
\caption{Conventional VAD Performance Benchmark, scored by ROC-AUC / F1@0.5}
\label{tab:vad-benchmark}
\setlength{\tabcolsep}{4pt}
\renewcommand{\arraystretch}{1.2}
\resizebox{\textwidth}{!}{%
\begin{tabular}{lccccccccccc}
\toprule
\multirow{2}{*}{\textbf{Model}} & \multirow{2}{*}{KAIST} & \multirow{2}{*}{Voxconverse} & \multirow{2}{*}{\makecell{Ten-VAD\\test}} & \multirow{2}{*}{\makecell{In-house}} & \multicolumn{7}{c}{LibriVAD concat} \\
\cmidrule(lr){6-12}
 & & & & & clean & SNR=-5 & SNR=0 & SNR=5 & SNR=10 & SNR=15 & SNR=20 \\
\midrule
webRTC$^1$                 & -- / 0.688 & -- / 0.433 & -- / 0.891 & -- / 0.587 & -- / 0.930 & -- / 0.288 & -- / 0.332 & -- / 0.332 & -- / 0.465 & -- / 0.564 & -- / 0.675  \\
Auditok$^2$                & -- / 0.586 & -- / 0.941 & -- / 0.886 & -- / 0.614 & -- / 0.938 & -- / 0.802 & -- / 0.802 & -- / 0.803 & -- / 0.807 & -- / 0.834 & -- / 0.899 \\
SpeechBrain\cite{speechbrain}            & 0.966 / 0.831 & 0.858 / 0.960 & 0.677 / 0.840 & 0.842 / 0.560 & 0.874 / 0.916 & 0.808 / 0.870 & 0.823 / 0.886 & 0.821 / 0.893 & 0.816 / 0.896 & 0.811 / 0.898 & 0.808 / 0.899  \\
Pyannote VAD\cite{bredin2019pyannoteaudioneuralbuildingblocks}           & -- / 0.945 & \textbf{-- / 0.975} & -- / 0.917 & -- / 0.655 & -- / 0.950 & -- / 0.890 & -- / 0.909 & -- / 0.917 & -- / 0.923 & -- / 0.928 & -- / 0.934  \\
FSMN-VAD\cite{FunASR}  & -- / 0.932 & -- / 0.972 & -- / 0.871 & -- / 0.705 & -- / 0.918 & -- / 0.847 & -- / 0.895 & -- / 0.901 & -- / 0.902 & -- / 0.903 & -- / 0.904  \\
Ten VAD~\cite{TENVAD}   & 0.989 / 0.927 & 0.930 / 0.949 & 0.942 / 0.928 & 0.969 / 0.851 & 0.980 / 0.955 & 0.835 / 0.807 & 0.882 / 0.880 & 0.912 / 0.904 & 0.932 / 0.918 & 0.948 / 0.927 & 0.960 / 0.934  \\
MarbleNet~\cite{jia2021marblenetdeep1dtimechannel}              & \textbf{0.994 / 0.947} & 0.962 / 0.966 & 0.920 / 0.912 & 0.973 / 0.616 & 0.979 / 0.955 & \textbf{0.890 / 0.895} & 0.919 / 0.912 & 0.935 / 0.919 & 0.945 / 0.925 & 0.953 / 0.929 & 0.958 / 0.933  \\
Silero-v5\cite{SileroVAD}              & 0.992 / 0.926 & 0.947 / 0.946 & 0.925 / 0.903 & 0.966 / 0.728 & 0.979 / 0.952 & 0.846 / 0.809 & 0.909 / 0.905 & 0.943 / 0.922 & 0.963 / 0.932 & 0.971 / 0.941 & 0.975 / 0.949  \\
Silero-v6\cite{SileroVAD}              & 0.992 / 0.947 & 0.952 / 0.952 & \textbf{0.957 / 0.939} & 0.966 / 0.862 & \textbf{0.981} / \textbf{0.960} & 0.846 / 0.831 & \textbf{0.918 / 0.914} & 0.946 / 0.930 & 0.963 / 0.937 & 0.972 / 0.944 & \textbf{0.976} / 0.953  \\
\textbf{Mamba-FVAD (IA)} & 0.986 / 0.910 & 0.933 / 0.929 & 0.910 / 0.890 & \textbf{0.982 / 0.884} & 0.979 / \textbf{0.960} & 0.830 / 0.712 & 0.916 / 0.884 & \textbf{0.958 / 0.933} & \textbf{0.970 / 0.947} & \textbf{0.974 / 0.953} & \textbf{0.976 / 0.956} \\
\bottomrule
\end{tabular}%
}
\end{table*}

In VOiCES the distractors form a built-in control: music is non-speech, so its BG-FAR measures only generic non-speech suppression, whereas babble and telephone are competing speech. The gap between babble/telephone and music thus isolates false alarms specific to background speech; a true speaker-rejecter stays near its music floor, a generic ``any-speech'' detector does not. Mamba-FVAD behaves like the former: telephone (0.07) sits at the music floor (0.06) and babble (0.12) only just above, a small speech-specific excess, while Silero shows a much larger gap (music 0.12 vs.\ telephone 0.18, babble 0.27). On this out-of-distribution distant-speech set Mamba-FVAD pays a small recall penalty (lower Foreground F1) yet rejects background speech as well as or better than the enrolled pVAD---without enrollment.


Overall, varying the backbone (Mamba$\to$LSTM) under the IA recipe largely preserves the selective behavior, whereas varying the recipe (IA$\to$LibriVAD) collapses it to that of a conventional VAD. We thus attribute foreground selectivity to interference-aware training, not architectural inductive bias.

\subsection{Conventional VAD Performance}

Foreground selectivity must not cost ordinary detection: with no competing speaker, an FVAD model should reduce to a conventional VAD rather than suppress single-talker speech. We verify this on three public sets: KAIST$^3$, VoxConverse~\cite{chung2020spot}, Ten-VAD test$^4$; and two deployment-relevant conditions: a human-annotated in-house benchmark and long-form LibriVAD-concat. The in-house set is 9.8\,h of conversational Japanese from real voice-agent interactions in noisy venues (restaurants, meeting rooms, open-plan offices) where background speech and babble are prevalent. Following FVAD, frame-level human labels mark only the intended speaker as speech, whereas background/overlapping talkers and non-speech are negative. Because such calls interleave clean single-talker and competing-speech stretches, it jointly tests conventional detection and foreground selectivity under one label set. For LibriVAD-concat we follow~\cite{stylianou2025librivadscalableopendataset}, concatenating LibriSpeech recordings under clean and additive-noise (SNR $-5$ to $20$\,dB) conditions. As not all baselines expose per-frame posteriors, we report ROC-AUC where available and F1@$0.5$ threshold as a secondary measure (Table~\ref{tab:vad-benchmark}).
{
\let\thefootnote\relax
\footnote{$^1$https://github.com/wiseman/py-webrtcvad}
\footnote{$^2$https://github.com/amsehili/auditok}
\footnote{$^3$https://github.com/jtkim-kaist/VAD}
\footnote{$^4$https://github.com/TEN-framework/ten-vad/tree/main/testset}
}



On the public sets, Mamba-FVAD is broadly competitive, trailing the strongest specialist on each (MarbleNet on KAIST, Pyannote on VoxConverse, Silero-v6 on TEN-VAD) only slightly and with no categorical failure, while leading on the in-house benchmark, whose labeling mirrors the FVAD objective. On LibriVAD-concat it is strong from $5$\,dB SNR upward, reflecting the conventional noise/music/RIR augmentation the IA recipe includes for robustness. The one weak spot is $-5$\,dB speech-shaped noise (SSN: ROC-AUC 0.70, vs $\geq$0.93 for non-speech noise). This represents the boundary case of the selectivity prior, whose training interferers are always quieter than the target, so a louder, sustained speech-like masker inverts the dominance cue and the model suppresses the now-quietest true foreground.

In sum, Mamba-FVAD trades a small, localized amount of generic-VAD accuracy for the strong foreground selectivity of Section~\ref{sec:bgfar}: it leads on the real-world in-house benchmark, leads from 5\,dB SNR upward on LibriVAD-concat, and stays within a few points of specialist VADs elsewhere. Because the IA recipe pairs competing-speech mixing with standard noise augmentation, selectivity and general noise-robustness coexist---the speaker-selective objective adds the former without sacrificing the latter.



\subsection{Ablations}
\label{sec:ablations}
For ablation experiments, all models were trained on the same IA recipe as Mamba-FVAD but on a fixed budget limited to 5 epochs. 


\textbf{Competing-speaker mixing.} Table~\ref{tab:overlap-arm} isolates the selectivity-bearing augmentation via three variants that differ only in how overlapping speech is generated: the reference (\emph{On}) renders each interferer as far-field background; \emph{Off} completely removes competing-speaker mixing; \emph{Nearfield} keeps the overlap but skips far-field simulation. \emph{Off} is by far the most damaging. BG-FAR rises across the board, widening as the interferer grows louder and largest on the real-recorded VOiCES. Never exposed to competing speech, the model tends to fire on any speech (hence its marginally higher foreground recall). This confirms that supervised exposure to \emph{unlabeled} competing speech, not generic noise augmentation, is the causal ingredient for background-speech rejection.
\begin{table}[h]
\centering
\caption{Competing-speaker augmentation ablation: BG-FAR and foreground recall (fgRec) on Mix-Interference and VOiCES.}
\label{tab:overlap-arm}
\setlength{\tabcolsep}{4pt}
\renewcommand{\arraystretch}{1.2}
\resizebox{\columnwidth}{!}{%
\begin{tabular}{lcccccccccc}
\toprule
\multirow{3}{*}{\makecell{Speaker\\mixing\\}} & \multicolumn{6}{c}{Mix-Interference} & \multicolumn{4}{c}{VOiCES} \\
\cmidrule(lr){2-7} \cmidrule(lr){8-11}
 & \multicolumn{5}{c}{BG-FAR $\downarrow$} & fgRec $\uparrow$ & \multicolumn{3}{c}{BG-FAR $\downarrow$} & fgRec $\uparrow$ \\
\cmidrule(lr){2-6} \cmidrule(lr){7-7} \cmidrule(lr){8-10} \cmidrule(lr){11-11}
 & S17 & S15 & S13 & S11 & S9\,(loud) & S9 & musi & babb & tele & babb \\
\midrule
On (ref)   & \textbf{0.059} & \textbf{0.088} & 0.140 & 0.226 & \textbf{0.357} & 0.946 & \textbf{0.141} & \textbf{0.270} & \textbf{0.183} & 0.848 \\
Off        & 0.076 & 0.112 & 0.184 & 0.300 & 0.462 & 0.957 & 0.231 & 0.342 & 0.235 & \textbf{0.877} \\
Nearfield  & 0.065 & 0.091 & \textbf{0.134} & \textbf{0.219} & 0.359 & \textbf{0.962} & 0.193 & 0.313 & 0.230 & 0.858 \\
\bottomrule
\end{tabular}%
}
\end{table}

\emph{Nearfield} shows that how the overlap is rendered governs real-world transfer. On synthetic Mix-Interference near- and far-field are nearly indistinguishable, but on VOiCES, where competitors are physically distant and reverberant, the far-field reference wins at every distractor, its training interferers matching real background acoustics. For only a one to two point recall cost, far-field competing-speaker mixing both rejects background speech and generalizes.



\textbf{Architecture.} Table~\ref{tab:arch-ablations} compares the three iso-parameter temporal decoder backbones under the identical IA recipe on the conventional VAD benchmarks. Mamba only very slightly edges out the LSTM, differing by at most $\sim$2 points of ROC-AUC, with each leading on some sets (Mamba on KAIST 0.962 and TEN-VAD 0.910, the LSTM on VoxConverse 0.930), at comparable parameter count. The Transformer trails on every benchmark despite its larger context. However, we acknowledge that the Transformer trained less stably on this data regime despite hyperparameter tuning and also required chunking on long clips; hence we treat its scores as a floor.


\begin{table}[h]
\centering
\caption{Architecture ablation: ROC-AUC across benchmarks. LibriVAD-concat is frame-weighted over all noises and SNRs.}
\label{tab:arch-ablations}
\setlength{\tabcolsep}{3pt}
\renewcommand{\arraystretch}{1.1}
\resizebox{\columnwidth}{!}{%
\begin{tabular}{lccccc|c}
\toprule
\textbf{Decoder} & KAIST & \makecell{Vox\\conv.} & \makecell{Ten\\VAD} & \makecell{In-\\house} & \makecell{LibriVAD\\concat} & \textbf{Params} \\
\midrule
Transf.  & 0.893 & 0.746 & 0.858 & 0.927 & 0.876 & 615{,}369 \\
LSTM     & 0.953 & \textbf{0.930} & 0.891 & 0.977 & 0.922 & 662{,}657 \\
Mamba    & \textbf{0.962} & 0.922 & \textbf{0.910} & \textbf{0.978} & \textbf{0.923} & 615{,}297 \\
\bottomrule
\end{tabular}%
}
\end{table}

Together with the matched Mamba/LSTM selectivity in Section~\ref{sec:bgfar} and the competing-speaker ablation in Section~\ref{sec:ablations}, the results suggest that foreground selectivity is driven primarily by the proposed supervision strategy, while the choice of backbone appears to play a secondary role under the evaluated settings. The backbone is therefore a deployment choice: we adopt Mamba for its $\mathcal{O}(1)$ streaming recurrence, with the LSTM a near-equivalent, lighter alternative.

\subsection{Performance Analysis}

\begin{figure}[htbp]
    \centering
    \includegraphics[width=1.0\columnwidth]{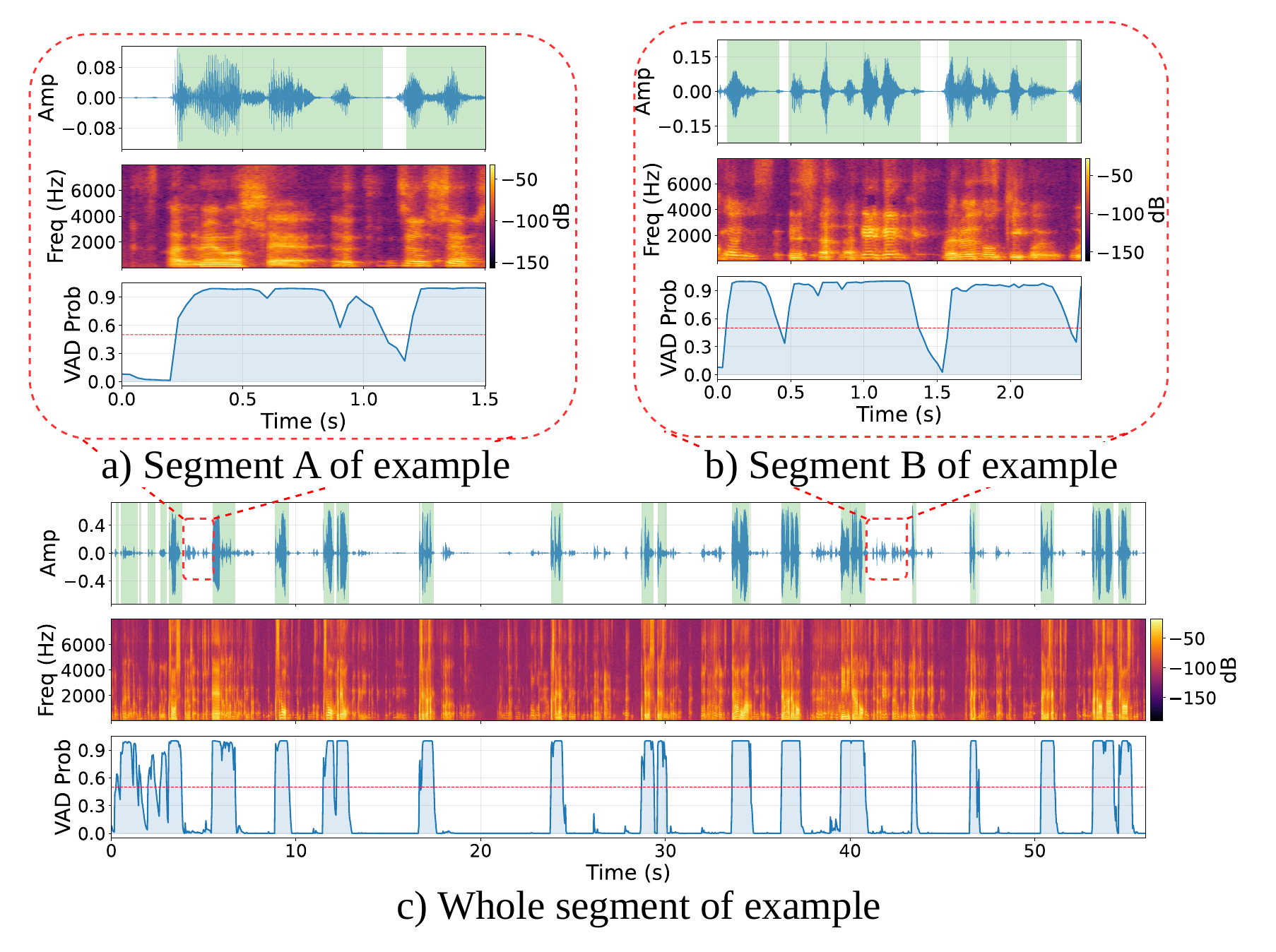}
    \caption{Mamba-FVAD inference on a multi-talker recording: the model locks onto the dominant speaker (c) while gracefully degrading to standard VAD in the absence of a primary speaker (a, b).}
    \label{fig:infer_example}
\end{figure}



To visualize the learned selectivity qualitatively, we overlay Mamba-FVAD's frame-level predictions on the input waveform for representative multi-talker clips (Fig.\ref{fig:infer_example}). The model's behavior tracks whether a foreground speaker has been established. Before the primary speaker enters (first $\sim$3\,s of Fig.\ref{fig:infer_example}c), no dominant speaker yet exists, so the model falls back to conventional VAD and marks the preceding speech-plus-noise as active. Once the primary speaker appears, it locks onto that target, and when the speaker stops ($\sim$4\,s later) it returns to silence even though background speech continues. Because the background spectrum is nearly identical before and after the target's turn, this switch cannot be explained by acoustic change alone; it reflects learned contextual selection.

Selection is not driven by loudness. Near the end of an utterance ($\sim$31\,s), the target is quiet yet still detected, indicating the model holds a stable latent representation of the foreground rather than tracking instantaneous amplitude. Background talkers are consistently suppressed with clean segment boundaries, and because FVAD carries no explicit speaker-embedding module, it follows speaker changes without re-enrollment, keeping the model lightweight. When no primary speaker is present (Figs.~\ref{fig:infer_example}(a),(b)), it gracefully reverts to conventional VAD, marking all speech active. Overall, Mamba-FVAD behaves like human auditory attention---focusing on a target, suppressing competitors, and falling back to general-purpose VAD when no target is established---at $1$--$2$\,ms per-frame latency on an AWS t2.micro instance, confirming real-time CPU-only deployment.

\section{Conclusion and Limitations}



We formalized Foreground VAD (FVAD), an enrollment-free, frame-synchronous task that tracks a single dominant speaker and reduces to conventional VAD when only one is present, and introduced BG-FAR and Foreground F1 to measure it. Our central finding is that foreground selectivity primarily comes from supervision: a fully automatic recipe pairing foreground-only labels with competing-speaker mixing turns an ordinary streaming VAD selective, and iso-parameter Mamba, LSTM, and Transformer backbones confirm that temporal-modeling capacity is at most a secondary factor. The resulting model, Mamba-FVAD, surpasses commercial VADs and enrollment-based speaker-aware systems on selectivity while staying competitive on conventional VAD and maintaining $1$--$2$\,ms per-frame CPU latency. The Mix-Interference benchmark will be released to support further study.

The selectivity that drives these gains also bounds the method. Because the model commits to a dominant foreground, it under-fires when the target is not the dominant source: recall falls at extreme negative SNR on all-speech-positive benchmarks, and on out-of-distribution far-field speech (VOiCES) selectivity transfers but foreground recall drops under the acoustic shift. We also assume a single foreground speaker per segment, leaving turn-taking and co-equal speakers to future work.

\newpage
\bibliographystyle{IEEEtran}
\bibliography{references}

\section*{Appendix A: Foreground F1 algorithm}
\label{ff1_algo}

\begin{algorithm}[h]
\caption{Foreground F1. Positives are foreground-speech frames; precision penalizes any
  activation on a foreground-silent frame (silence, noise, \emph{or} competing speech),
  recall penalizes missed foreground speech.}
\label{alg:fgf1}
\KwIn{model $f$; threshold $\theta=0.5$; clips $(x,\mathbf{y})$ with foreground labels $\mathbf{y}$}
\KwOut{Foreground F1}
$\mathrm{TP}\leftarrow 0$,\ \ $\mathrm{FP}\leftarrow 0$,\ \ $\mathrm{FN}\leftarrow 0$\;
\ForEach{clip $(x,\mathbf{y})$}{
  $\mathbf{p} \leftarrow f(x)$\tcp*{per-frame foreground posteriors}
  \For{$t \leftarrow 1$ \KwTo $T$}{
    $\hat{y}_t \leftarrow [\,p_t \ge \theta\,]$\tcp*{$1$ if $p_t\ge\theta$, else $0$}
    \uIf{$\hat{y}_t = 1$ \textbf{and} $y_t = 1$}{$\mathrm{TP} \leftarrow \mathrm{TP}+1$}
    \uElseIf{$\hat{y}_t = 1$ \textbf{and} $y_t = 0$}{$\mathrm{FP} \leftarrow \mathrm{FP}+1$}
    \ElseIf{$\hat{y}_t = 0$ \textbf{and} $y_t = 1$}{$\mathrm{FN} \leftarrow \mathrm{FN}+1$}
  }
}
$P \leftarrow \mathrm{TP}/(\mathrm{TP}+\mathrm{FP})$,\ \ $R \leftarrow \mathrm{TP}/(\mathrm{TP}+\mathrm{FN})$\;
\Return $2PR/(P+R)$\;
\end{algorithm}

\FloatBarrier

\clearpage
\section*{Appendix B: BG-FAR algorithm}
\label{bgfar_algo}

\begin{algorithm}[h]
\caption{Background False-Alarm Rate (BG-FAR).
  $\textsc{LogEnergy}$ returns per-frame log-energy
  $E_t = 10\log_{10}(\tfrac{1}{H}\sum_{i=1}^{H} x_{(t-1)H+i}^{2})$ on the
  31.25\,Hz grid ($H{=}512$); the reference $x^r$ is the \texttt{noise\_only}
  take (Mix-Interference) or the \texttt{none} take (VOiCES).}
\label{alg:bgfar}
\KwIn{model $f$; threshold $\theta=0.5$; margin $\delta=6$\,dB;
      clips paired as $(x^c,x^r,\mathbf{y})$ -- condition $x^c$ (foreground + competing speaker),
      matched reference $x^r$ (same scene, no competing speaker), foreground labels $\mathbf{y}$}
\KwOut{BG-FAR}
$n \leftarrow 0$\tcp*{false alarms on background-active frames}
$d \leftarrow 0$\tcp*{total background-active frames}
\ForEach{paired clip $(x^c,x^r,\mathbf{y})$}{
  $\mathbf{p} \leftarrow f(x^c)$\tcp*{per-frame foreground posteriors}
  $E^c \leftarrow \textsc{LogEnergy}(x^c)$,\ \ $E^r \leftarrow \textsc{LogEnergy}(x^r)$\;
  \For{$t \leftarrow 1$ \KwTo $T$}{
    \If{$y_t = 0$ \textbf{and} $E^c_t - E^r_t > \delta$}{
      $d \leftarrow d + 1$\;
      \lIf{$p_t \ge \theta$}{$n \leftarrow n + 1$}
    }
  }
}
\Return $n/d$\;
\end{algorithm}

\FloatBarrier

\section*{Generative AI Use Disclosure}
Generative AI tools (ChatGPT, Claude) were used for grammar checking and polishing the English writing of this manuscript. All technical content, experimental design, and scientific conclusions were produced entirely by the authors.

\end{document}